\documentclass[aps,prl,superscriptaddress,twocolumn,floatfix,nofootinbib]{revtex4}
\usepackage{graphicx}  
\usepackage{dcolumn}   
\usepackage{bm}        
\usepackage{amssymb}   
\usepackage{amsmath}
\usepackage{epsfig}
\usepackage{rotating}
\usepackage{color}
\newcommand{\be}{\begin{equation}}
\newcommand{\ee}{\end{equation}}
\newcommand{\Fc}{{\cal F}}
\begin{document}


\title{ Geodesic completeness and homogeneity condition for cosmic inflation}

\author{Aindri\'u Conroy}
\affiliation{Consortium for Fundamental Physics, Physics Department, Lancaster  University, Lancaster, LA1 4YB, UK}

\author{Alexey S. Koshelev}
\affiliation{Theoretische Natuurkunde, Vrije Universiteit Brussel and The 
International Solvay Institutes, Pleinlaan 2, B-1050, Brussels, Belgium}

\author{Anupam Mazumdar}
\affiliation{Consortium for Fundamental Physics, Physics Department, Lancaster  University, Lancaster, LA1 4YB, UK}

\begin{abstract}
There are two disjointed problems in cosmology within General Relativity (GR), which can be addressed simultaneously by studying the nature of 
geodesics around $t\rightarrow 0$, where $t$ is the physical time. One is related to the past geodesic completeness of the inflationary trajectory due 
to the presence of a cosmological singularity,  and the other one is related to the homogeneity condition required to inflate a local space-time patch of 
the universe. We will show that both the problems have a common origin, arising from how the causal structure of null and timelike geodesics are structured within 
GR. In particular, we will show how a non-local extension of GR can address both problems, while satisfying the 
{\it null} energy condition for the matter sources.
\end{abstract}

\maketitle

Primordial inflation is extremely successful in explaining the current observed universe~\cite{Planck}. However, there are many fundamental issues with inflation. Two of the most important ones are related to its embedding within General 
Relativity (GR)~\footnote{Inflation has many other challenges, 
see~\cite{Linde:2005ht,Mazumdar:2010sa}.}.

\begin{itemize}
\item
{\textit{Geodesic incompleteness}: Due to the inevitability of a cosmological 
singularity within GR, inflationary trajectories are past-incomplete~\cite{Guth}. 
One can see that this warrants a better theory of gravity in the ultraviolet (UV), which would ameliorate the UV divergences 
as well as make the theory {\it singularity-free} in the 
UV, for instance~\cite{Biswas:2011gr,Biswas:2005qr,Tomboulis,Modesto}.
Such a singularity-free universe would yield a non-singular bouncing cosmology, and possibly this would leave some falsifiable
imprints in the sky~\cite{Biswas:2013dry}.}

\item
{\textit{Homogeneity condition}: Slow roll inflation requires a patch of 
the universe to be sufficiently homogeneous on super-Hubble scales, 
see~\cite{Linde:1993xx,Vachaspati:1998dy}, see also~\cite{Albrecht:1985yf,Albrecht}. In this respect, inflation
within GR does not solve the homogeneity problem - it assumes homogeneity to begin with. Even if inflation begins at the Planckian epoch, one requires the spatial 
gradient terms in the action of the inflaton field (whose slow roll leads to inflation) to be sufficiently negligible compared to the homogeneous, time dependent terms. }

\end{itemize}

A priori, these two problems seem to be unrelated. However, they have a common 
origin and if  the first one is addressed, then the second one can also be 
understood, 
which would lead to a better understanding of inflation within a UV complete theory of gravity~\cite{Chialva:2014rla}. They are both related to the causal structure of the spacetime
within GR, assuming the weak energy condition (WEC) for the matter field, i.e. $\rho\geq0$ and $\rho+p \geq 0$, which necessarily implies the null energy condition (NEC), 
$\rho+p\geq0$, where $\rho$ is the energy density and $p$ is the pressure component.

The main aim of this paper is to build this connection and show how a geodesically past-complete universe would naturally evade the constraints of the
homogeneity condition for slow roll models of inflation. We will illustrate this problem by modifying the UV aspects of gravity and therefore modifying
the causal structure of the spacetime. In particular, we will  invoke a
non-local modification of GR, which can be made ghost-free in the UV, while also recovering GR and its predictions in the infrared (IR)~\cite{Biswas:2011gr,Biswas:2005qr}.

{\bf Causal structure of spacetime and the Raychaudhuri Equation:} The structure of a singularity can be understood in a {\it model independent} way by studying 
the Raychaudhuri Equation (RE) for timelike and/or null geodesic congruences. For simplicity, we consider only null geodesic congruences such that $k^{\mu}k_{\mu}=0$, where $k_{\mu}$ is a four vector tangential to the null geodesic congruence, defined by mostly positive convention, i.e. $(-,+,+,+)$, and the expansion parameter, $\theta$, defined by $\theta= \nabla_{\mu}k^{\mu}$. Let us concentrate on the simplest possible scenario where the \emph{twist} tensors vanish, which is true
if we take the congruence of null rays to be orthogonal to the hypersurface. 
Furthermore, the shear tensor is purely spatial and thus makes a positive 
contribution. Taking all of this into account the RE can be 
simplified greatly, see~\cite{Wald}
\be\label{eq1}
 \frac{d\theta}{d\tau}+\frac{1}{2}\theta^{2}\le-R_{\mu\nu}k^{\mu}k^{\nu}
 \ee
where $\tau$ is the affine parameter,  and $R_{\mu\nu}$ 
is the Ricci tensor. 

We know from the Einstein equation that $G_{\mu\nu}=\kappa T_{\mu\nu}$, where $\kappa =8\pi G=M_{p}^{-2}$, which in turn implies $R_{\mu\nu}=\kappa( T_{\mu\nu}-\frac{1}{2} g_{\mu\nu} T)$. Now, contracting with the vector field $k^{\mu}$, we find $R_{\mu\nu}k^{\mu}k^{\nu}=\kappa T_{\mu\nu}k^{\mu}k^{\nu}$. Finally, imposing the NEC, $T_{\mu\nu}k^{\mu}k^{\nu}\geq 0$, 
we obtain the \emph{null convergence condition} (null CC) expressed in two equivalent ways:
 \be\label{eq2}
 R_{\mu\nu}k^{\mu}k^{\nu}\geq 0,\qquad
 \frac{d\theta}{d\tau}+\frac{1}{2}\theta^{2}\le0
 \ee
 This suggests that the converging null geodesics cannot start to diverge before 
meeting the 
{\it origin of coordinates} or, in other words, the converging null geodesic must meet the space-like singularity 
in a finite time within GR, where the NEC is satisfied~\cite{Borde:1996pt, Borde}.

 {\bf Trapped, antitrapped and normal surfaces:} In an asymptotically flat spacetime, \emph{trapped} surfaces and an apparent horizon are formed when when both the ingoing and outgoing expansions are negative, i.e. $\theta_{IN,OUT}< 0$. A period of cosmic acceleration with positive ingoing and outgoing expansion, $\theta_{IN,OUT}>0$, gives rise to \emph{antitrapped} surfaces and \emph{normal} regions are defined by the behaviour $\theta_{IN}<0$ and $\theta_{OUT}>0$. Any surface with physical size greater or equal to the \emph{minimally antitrapped surface} (MAS) has \emph{vanishing expansion} and is, by definition, antitrapped. Within the Friedmann-Robertson-Walker (FRW) metric, $ds^2=dt^2-a^2(t)dr^2$, where $a(t)$ is the scale factor and $r$ is the coordinate of $3$ spatial directions, $x_{MAS}=H^{-1}(t)$, where $H(t)\equiv \dot a(t)/a(t)$ and the physical size of the MAS is represented by $x_{MAS}$.
 The inner boundary of such a surface is known as the \emph{cosmological apparent horizon} and is defined as the inverse of the physical distance of the MAS of the background cosmology such that
\be
x_{FRW}=x_{MAS}=H^{-1}_{FRW}
\ee 
Similarly, in Fig.1 the line OQ denotes the inflationary patch, with the segment OP equal to the inverse of the inflationary apparent horizon which is necessarily of smaller physical size to the MAS. Now, in the usual  FRW universe - complete with an initial singularity by virtue of the NEC condition within GR, see Eq.~(\ref{eq2}) - the ingoing null ray cannot go from a normal region to an antitrapped region, as depicted by the arrow. As argued in \cite{Vachaspati:1998dy}, the inflationary patch must be {\it embedded} already within an antitrapped region of spacetime in order to trigger inflation without violating the null CC. The conclusion was that late inflation requires a prior phase of inflation. However with an impending singularity in the past, one would be left with inflation occurring already at the 
Planckian epoch~\cite{Linde:1993xx,Vachaspati:1998dy}.

\begin{figure}
\centering
\includegraphics[width=45mm,height=50mm]{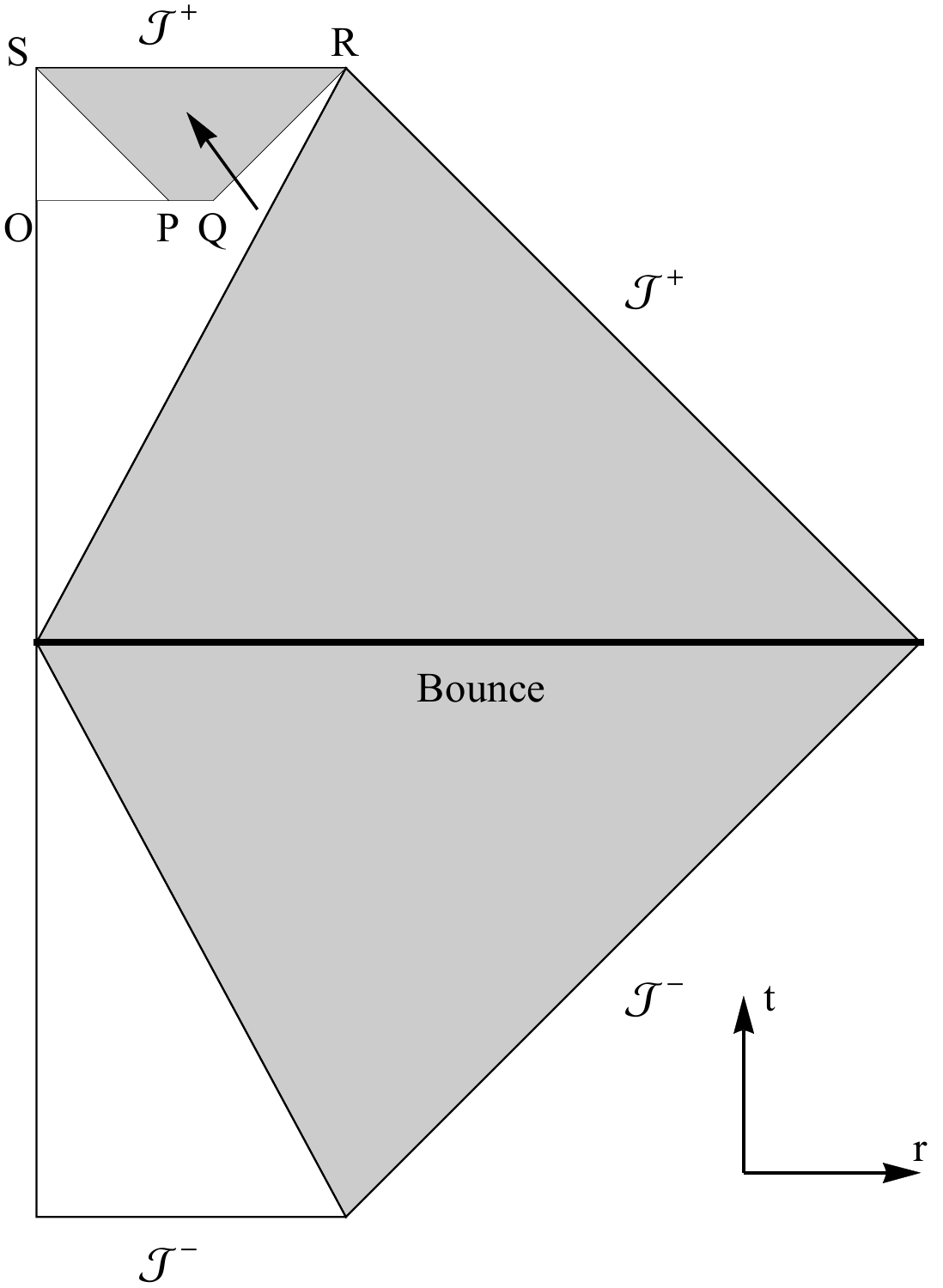}
\caption{A conformal diagram depicting a bouncing cosmology in an isotropic and 
homogenous FRW metric. Shaded regions are antitrapped and unshaded regions are 
normal. A patch begins to inflate at later times, $t$, from O to Q with 
inflationary size $x_{inf}$, where the line OP borders the apparent inflationary 
horizon. The arrow, pictured, describes an ingoing null ray entering an 
antitrapped surface from a normal surface. All notations are as in 
\cite{Vachaspati:1998dy}.}
\end{figure} \label{fig}

{\bf Non-singular bouncing cosmology:}  
 As we shall see below, a non-singular bouncing cosmology naturally leads to an accelerated expansion near the bounce, $\ddot a(t)> 0$. The challenge 
 is to realise a non-singular bounce which requires modification to GR. In particular, a reversal of the inequality in the null CC without violating the 
 NEC, would allow the converging null rays to be made past complete, thus resolving {\it not only} the cosmological singularity problem, but also 
 allowing the arrow shown in Fig.1 to go from a normal region of spacetime to an antitrapped region of spacetime. Therefore, the {\it homogeneity} 
 condition for inflation is ameliorated, especially at later stages. 
 
 In a word, a non-singular bouncing cosmology  naturally provides all the necessary conditions for successful inflation,
 which must occur at later stages in order to produce the large scale structures present in the universe.

{\bf Modifying GR in the UV:} We may now ask the question: 
what modification of GR would yield a reversal of the inequality contained within the null CC, such that 
\be\label{eq3}
 R_{\mu\nu}k^{\mu}k^{\nu}\leq 0,\qquad
 \frac{d\theta}{d\tau}+\frac{1}{2}\theta^{2}\geq 0
 \ee
and thus describe a singularity-free theory of gravity, whilst retaining the NEC? There are two generic ways in which this may be satisfied.

\begin{itemize}

\item
{\textit{Local modification of GR}:  Higher order corrections such as 
${\cal L}\sim c_2R^2+c_3R^3+\cdots +d_2R_{\mu\nu}^2, \cdots + 
e_2R_{\mu\nu\lambda\sigma}^2,\cdots$, 
with appropriate coefficients would modify the UV
behaviour of gravity. The higher derivatives help to ameliorate the UV aspects of gravity in $4$ dimensions but they typically contain {\it ghosts}. This has been known from the days of
Stelle's theory of $4th$ order gravity, which is renormalizable but contains massive ghosts~\cite{Stelle}.}

\item
{\textit{Non-local modification of GR}: The ghost problem can be 
addressed in the case of {\it infinite} higher order derivatives. Let us 
concentrate on quadratic curvature
with ${\cal L}\sim R{\cal F}(\Box)R+ R_{\mu\nu}{\cal G}(\Box)R^{\mu\nu}$, where ${\cal F}(\Box),~{\cal G}(\Box)$ are {\it analytic functions} containing higher derivatives up to infinite order, where $\Box =g^{\mu\nu}\nabla_{\mu}\nabla_{\nu}$
is the d'Alembertian operator. In the Minkowski background, these comprise the most generalised action of gravity with non-local contributions, yielding a
{\it ghost-free} condition for certain analytic choices of ${\cal F,~G}$, 
constructed, necessarily, from an {\it entire 
function}~\cite{Biswas:2005qr,Biswas:2011gr,Tomboulis,Modesto}.}

\end{itemize}

{\bf Explicit Example:} In order to illustrate and for the remainder of the 
paper, let us concentrate on a non-local generalisation of 
GR in the UV, up to quadratic order in curvature,
\be
S=\int d^4x\sqrt{-g}\biggl(\frac{M_P^2}{2}R+\frac{R{\cal 
F}(\Box)R}2\biggr)
\label{action}
\ee
where $
{\cal F}(\Box)\equiv\sum\limits^{\infty}_{n=0}\frac{f_{n}}{M_p^{2n}}\Box^n
$ and, without loss of generality, we have assumed that non-renormalisable 
operators are suppressed by the 
$4$ dimensional scale of gravity, i.e. $M_p$.

It is important to add, that the RE and its convergence conditions hold 
independently of the background action. However, we must introduce an action in 
order to compute the Ricci tensor. We compute the equations of motion as in  
\cite{Biswas:2013cha}, and impose the FRW metric.

In order to understand the nature of these null geodesic congruences, let us 
concentrate on the simplest regime with a {\it homogeneous} metric
near $t=0$. To this end, we may describe the scale factor expanding around   
$t=0$, as follows
\be\label{at}
a(t)=1+a_2 t^2+{\cal O}(t^4)+\cdots\,.
\ee
Note that we may consider \emph{even} powers of $t$ to understand the 
solution around $t=0$,
with coefficient $a_2>0$ as a consequence of ${\ddot a}(t)>0$ at the bounce point, since we are interested in seeking a non-singular solution.
This implies that $H(t)$ is an \emph{odd} function of physical time $t$, while 
$R(t)=12H^2(t)+6{\dot H}(t)$ is an even function of $t$ such that
\be
\label{Rt}
R(t)=R_0+R_2t^2+{\cal O}(t^4)+\cdots\,.
\ee
with $R_0>0$ at the bounce, as $\lim_{t\rightarrow0}R(t)=12a_2$ and, as we have already stated, ${\ddot a}>0$ implies $a_2>0$ at the bounce point.

By solving for the equations of motion of the action given by Eq.~(\ref{action}) (for details, see~\cite{Biswas:2013cha}), one can 
extract the Ricci tensor $R_{\mu\nu}$ and contract with the vector field $k^\mu$ to find, at the bounce point $t=0$:
\be
\label{Ruvkukv}
R_{\mu\nu}k^\mu 
k^\nu=(k^0)^2\frac{(\rho+p)
+2\partial_t^2({
\cal F}(\Box)R)}
{M_p^2+2{\cal 
F}(\Box)R}.
\ee
Next, we must compute the non-local terms ${\cal F}(\Box)R$ and 
$\partial_t^2{\cal F}(\Box)R$, Eqs. (\ref{F0},\ref{d2F0}). For detailed steps, 
see Appendix.

Crucially, we are now in a position to deduce that, in order for the l.h.s. of  Eq.~\eqref{Ruvkukv} to be negative, i.e. 
$R_{\mu\nu}k^\mu k^\nu\leq 0$, thus determining the conditions for which null geodesics can be made past-complete, 
the following inequalities must hold for either upper or lower signs,
\be
\label{inequalities}
\frac{(\rho+p)}{2R_0}\lessgtr y\Fc(y),
\quad \quad \quad\quad \quad  \frac{M_{ p }^{ 2 }}{2R_0}\gtrless 
-\Fc(y),
\ee 
where $y\equiv -2R_2/R_0$ is defined in the Appendix, and we have assumed the 
NEC to hold true always.\\  

{\bf Ghost-Free choice:}
Following Ref.~\cite{Biswas:2005qr,Biswas:2011gr},  a particular class of 
${\cal F}(\Box)$ can be chosen in order to make a non-local theory of 
gravity ghost-free without violating general covariance. Here, we choose~\cite{Biswas:2005qr,Biswas:2011gr}:
\be
\label{GF}
{\cal F}(\Box)=\frac{e^{-\Box/M_p^{2}}-1}{\Box/M_p^{2}}
\ee
Note that due to the particular nature of Eq.~(\ref{GF}), we have ${\cal 
F}(y<0)<-1$, and $-1\leq\Fc(y\geq 0)<0$.

In order to extract the physics, we may entertain the simplest interesting scenario when the {\it curvature } of the 
universe is evolving {\it adiabatically} near the Planck scale, in such a way that $R_2$ is small, i.e. $R_2\ll R_0$ in
Eq.~(\ref{Rt}). This is justifiable since we are ignoring the higher order terms 
 in both our expressions, Eqs.~(\ref{at},~\ref{Rt}).
We will be interested then in a limit when $y\rightarrow 0$. 
Note that at this point $R_2$ could be either
positive or negative.

\begin{itemize}

\item
{$R_2\geq 0$: when $y\rightarrow 0$,  the \emph{lower signs} in 
Eq. (\ref{inequalities})
yield: 
\be
 \label{cond20}
\frac{(\rho+p)}{2R_0}> 0,\quad\quad\quad\quad
~\frac{M_{ p }^{ 2 }}{2R_0}< 1\,.
 \ee
This tells us that the NEC must be satisfied, irrespective of $R_0$, and  
$R_0> M_p^2/2$ at the bounce.
Note that one would naturally expect $R_0\leq M_p^2$ at the time of the 
bounce.\\
The \emph{upper signs} do not yield any physically motivated solution when 
$R_2\geq 0$.
}

\item{$R_2\leq0$: The first inequality in Eq.~(\ref{inequalities}) holds true  
with the lower sign as long as $\rho+p>0$, yielding:
\be
\frac{M_P^2}{2R_0}<-\Fc(y)\leq1
\label{cond3}
\ee
which is analogous to Eq.~(\ref{cond20})}
\end{itemize}

For either  $R_2\geq 0$ or  $R_2\leq 0$,  one ensures that the null
geodesics are past-complete and a non-singular bouncing cosmology can be constructed near $t\sim 0$, at the limit of
$y\rightarrow 0$, without violating the NEC.

Past-completeness of null geodesics implies the past-completeness of timelike 
geodesics, independent of any choice of $\Fc(\Box)$, as can be shown by a straightforward 
computation. 

{\bf Discussion:} In this paper, we pointed out a neat solution for two disjointed problems of inflationary cosmology. 
We argued that a non-singular bouncing cosmology is a must for a successful inflationary paradigm, where one can 
make the inflationary trajectory past-complete, while also explaining the pre-requisite homogeneity condition for inflation by
modifying the Raychaudhuri equation. Thus, a reversal of the inequality in the 
null CC, see Eqs.~(\ref{eq2},~\ref{eq3}), allows both a non-singular 
bounce and the movement of null rays from a normal surface to 
antitrapped surface, without violating the NEC in the matter sector.

We showed that a non-local modification of GR would modify the Raychaudhuri 
equation in such a way that a bouncing 
non-singular cosmology can be constructed. In particular, we have shown that for
a  slowly varying curvature  with a scale, 
at bounce, comparable to the Planck scale within a homogeneous and 
isotropic metric, one can avoid a cosmological singularity. The matter at 
the same time enjoys the NEC.

Our 
results have 
implications for completing the past inflationary geodesics and also 
ameliorating the {\it homogeneity} conditions for inflation at later 
epochs. Note that a non-singular bounce naturally  provides a platform for an 
inflationary cosmology, by virtue of $\ddot a (t)> 0$,
at the bounce point, thus providing the initial conditions for primordial inflation. Similar analyses for avoiding cosmological singularity may be performed for a variety of modifications to GR in the UV.

{\it Acknowledgments:} AC is funded 
by STFC grant no ST/K50208X/1. AK is supported by an ``FWO-Vlaanderen'' 
postdoctoral fellowship and
also in part by the Belgian Federal Science
Policy Office through the Interuniversity Attraction Pole P7/37, by
FWO-Vlaanderen through project G011410N, by the Vrije Universiteit
Brussel through the Strategic Research Program ``High-Energy Physics'',
and by the RFBR grant 11-01-00894. AM is supported by the 
Lancaster-Manchester-Sheffield 
Consortium for Fundamental Physics under STFC grant ST/J000418/1.

 \section{Appendix}
Here we compute the non-local terms ${\cal F}(\Box)R$ and $\partial_t^2{\cal 
F}(\Box)R$.  We do this by first calculating $e^{s\Box}$, where $s$ is a 
constant. Using the diffusion equation method \cite{calcagni}, we find for 
$t\rightarrow0$
\be
(e^{s\Box}R)(0)=R_0 e^{sy}
\ee
where we have defined $y\equiv-2{R_{2}}/{R_{0}}$.
We then represent the operator ${\cal F}(\Box)$, using the inverse  Laplace 
integral transform, with the integration contour such that all poles are on one 
side of the contour with $\alpha$, real
\be
\Fc(\Box)=\frac 1{2\pi i}\int_{\alpha-i\infty}^{\alpha-i\infty} \tilde 
\Fc(s)e^{s\Box}ds\,,
\ee
The next step is to solve for the relevant non-local terms at the bounce. In the 
first instance, we find ${\cal F}(\Box)R$ at the bounce to be
\be\label{F0}
({\cal F}(\Box)R)(0)=R_{0}{\cal F}(y)
 \ee
 In order to compute, the second time derivative, we note that we are 
effectively computing $(-\Box{\cal F}(\Box)R)(0)$. The properties of the Laplace 
transform yield
\be\label{d2F0}
 (\partial_{t}^{2}{\cal F}(\Box)R)(0)=2R_{2}{\cal F}(y).
\ee


\begin{thebibliography}{99}
        
        
\bibitem{Planck}        
  P.~A.~R.~Ade {\it et al.}  [Planck Collaboration],
  Astron.\ Astrophys.\  (2014)
  [arXiv:1303.5076 [astro-ph.CO]].
  
 \bibitem{Linde:2005ht} 
  A.~D.~Linde,
  Contemp.\ Concepts Phys.\  {\bf 5}, 1 (1990)
  [hep-th/0503203].

\bibitem{Mazumdar:2010sa} 
  A.~Mazumdar and J.~Rocher,
  Phys.\ Rept.\  {\bf 497}, 85 (2011)
  [arXiv:1001.0993 [hep-ph]].

\bibitem{Guth}
A.~Borde, A.~H.~Guth and A.~Vilenkin,
  Phys.\ Rev.\ Lett.\  {\bf 90}, 151301 (2003)
  [gr-qc/0110012].
  
  
  \bibitem{Biswas:2011gr}
   T.~Biswas, E.~Gerwick, T.~Koivisto and A.~Mazumdar,
  Phys.\ Rev.\ Lett.\  {\bf 108}, 031101 (2012)
  [arXiv:1110.5249 [gr-qc]].
  
  
\bibitem{Biswas:2005qr}
  T.~Biswas, A.~Mazumdar and W.~Siegel,
  JCAP {\bf 0603} 009 (2006)
  [hep-th/0508194].
  
  \bibitem{Tomboulis}
  E. Tomboulis (1997), hep-th/9702146
  
  
  \bibitem{Modesto}
  L. Modesto, Phys.Rev. D86, 044005 (2012), 1107.2403,
  L.~Modesto, J.~W.~Moffat and P.~Nicolini,
  Phys.\ Lett.\ B {\bf 695}, 397 (2011).
  
  
  \bibitem{Biswas:2013dry} 
  T.~Biswas and A.~Mazumdar,
  Class.\ Quant.\ Grav.\  {\bf 31}, 025019 (2014)
  [arXiv:1304.3648 [hep-th]].
  
  
  \bibitem{Linde:1993xx} 
  A.~D.~Linde, D.~A.~Linde and A.~Mezhlumian,
  Phys.\ Rev.\ D {\bf 49}, 1783 (1994)
  [gr-qc/9306035].
        
\bibitem{Vachaspati:1998dy}
  T.~Vachaspati and M.~Trodden,
  Phys.\ Rev.\ D {\bf 61} (1999) 023502
  [gr-qc/9811037].
  
 \bibitem{Albrecht:1985yf}  
   A.~Albrecht, R.~H.~Brandenberger and R.~Matzner,
  Phys.\ Rev.\ D {\bf 32}, 1280 (1985).
  
  \bibitem{Albrecht}
  A.~Albrecht, R.~H.~Brandenberger and R.~Matzner,
  Phys.\ Rev.\ D {\bf 35}, 429 (1987).
  
 \bibitem{Chialva:2014rla} 
  D.~Chialva and A.~Mazumdar,
  (2014) arXiv:1405.0513 [hep-th].
  
  \bibitem{Wald}
  R. Wald, General Relativity (University of Chicago Press,
1984).
  
  \bibitem{Borde:1996pt} 
  A.~Borde and A.~Vilenkin,
  Int.\ J.\ Mod.\ Phys.\ D {\bf 5}, 813 (1996)
  [gr-qc/9612036].
  
  \bibitem{Borde}
   A.~Borde, A.~H.~Guth and A.~Vilenkin,
  Phys.\ Rev.\ Lett.\  {\bf 90}, 151301 (2003)
  [gr-qc/0110012].
  
  
  \bibitem{Stelle}
  K. Stelle, Phys.Rev. D{\bf 16}, 953 (1977)
 

  
\bibitem{Biswas:2013cha}
  T.~Biswas, A.~Conroy, A.~S.~Koshelev and A.~Mazumdar,
  Class.\ Quant.\ Grav.\  {\bf 31} 015022 (2014)
  [arXiv:1308.2319 [hep-th]].


\bibitem{calcagni}
G.~Calcagni, M.~Montobbio and G.~Nardelli,
Phys.\ Rev.\ D {\bf 76} 126001 (2007)
[arXiv:0705.3043 [hep-th]].

\end{thebibliography}
\end{document}